# GaN mid-IR plasmonics: low-loss epsilon-near-zero modes

*Julia Inglés-Cerrillo, Maria Villanueva-Blanco, Benjamin Damilano, Stéphane Vézian, Miguel Montes Bajo\*, Adrian Hierro\*.*

J. Inglés-Cerrillo, M. Villanueva-Blanco, M. Montes Bajo, A. Hierro
ISOM, Universidad Politécnica de Madrid, Spain
\* E-mail: miguel.montes@upm.es, adrian.hierro@upm.es

B. Damilano, S. Vézian
Université Côte d'Azur, CNRS, CRHEA, Valbonne, France

Epsilon-near-zero (ENZ) materials, defined by $|Re(\varepsilon)| < 1$, enable unique light propagation characteristics, including confinement within sub-wavelength regions. To reduce losses in this regime, materials with both near-zero permittivity and $n < 1$ refractive index, known as near-zero-index (NZI) materials, are desired. When both conditions are satisfied, the resulting region is classified as a low-loss ENZ medium combining strong light confinement with reduced optical losses. To achieve this behavior in the mid-IR, heavily doped semiconductors are required, and those compatible with current technologies are most desirable. This work provides the first in-depth study, supported by experimental demonstrations, of the plasmonic properties of highly doped GaN thin films on Si, exhibiting low optical losses and low-loss ENZ characteristics up to 3 μm. From the extracted optical parameters, the ENZ and NZI regions are determined and compared with the existing literature. As a result of the large polar character of nitrides, a hybridization of the surface plasmon and phonon polaritons is observed, accompanied by a flat-dispersion of the high-energy mode (pinned near the plasma frequency) indicative of its ENZ character. Establishing GaN as a viable platform for mid-IR ENZ-based plasmonics paves the way for integration into future infrared photonic technologies.

## 1. Introduction

Epsilon-near-zero (ENZ) materials, characterized by a real part of the dielectric permittivity approaching zero within a specific spectral range, have attracted significant attention in the field of plasmonics due to the extraordinary light phenomena they present.[1–3] These include the confinement of light within sub-wavelength thin films,[1,4] enhanced electric fields,[5,6] novel optical modes,[7,8] high emission directionality,[9–11] photon tunnelling,[12] or extreme non-linear interactions.[13–15] The ENZ regime is defined by the condition $|Re(\varepsilon)| < 1$.[1] This definition only takes into account the real part of the permittivity and therefore does not consider the optical losses. To overcome this limitation, achieving materials with not only a vanishing permittivity but also a vanishing refractive index is desirable as a way of minimizing the losses in the ENZ regime. The near-zero index (NZI) condition is defined as $n < 1$, and when both the ENZ and the NZI conditions are satisfied, the material can be considered a low-loss ENZ material.[1]

In the visible and near-IR regions, metals are the optimal choice to exploit the ENZ character near the electron plasma resonance, owing to their intrinsically high plasma frequencies that typically lie in the UV range. However, they become ineffective in the mid-IR and beyond, since their electron concentration cannot be tuned, and shifting their plasma resonance into the mid-IR requires the fabrication of nanostructures involving complex nanofabrication processes.[16–18] Nevertheless, the mid-IR spectral range is highly attractive and has been the subject of extensive research for many decades due to its broad range of applications, including molecular sensing, thermal imaging, and spectroscopy.[16] This continued interest has driven an ongoing search for materials exhibiting ENZ characteristics in the mid-IR, among which semiconductors have emerged as promising candidates.[19,20] Their ENZ character has been exploited through two approaches: the plasmonic approach, in which semiconductor doping allows tuning the plasma frequency across the frequency range of interest; and the phononic approach, which uses phonon resonances that naturally fall in the mid-IR but lack frequency tunability. In the plasmonic approach, where ENZ is a result of the carriers' response to an electric field oscillation in the electromagnetic waves,[2] there are several studies which demonstrate and exploit the ENZ character of different doped semiconductors such as GaAs,[21] GaInAs,[22] InGaAlAs,[23] InAsSb[24] and transparent conductive oxides (TCOs), such as CdO and ZnO.[25,26] Within the phononic approach, the ENZ response originates from the zero-crossing of the real part of the permittivity near the LO phonon resonance, and numerous works have shown the potential of this ENZ regime obtained with the phonon resonances of different polar semiconductors including SiC,[9] InP[27] or AlN.[28]

Gallium nitride (GaN) is one of the key semiconductors in industry today; yet, surprisingly, its potential in ENZ applications has only been scarcely studied. GaN stands out among all semiconductors since it offers three key advantages: it can be heavily doped (up to values of $N = 6.7 \times 10^{20}$ cm$^{-3}$ using Ge as n-type dopant),[29] enabling a large control of its plasma frequency; it is highly polar, featuring a Reststrahlen band of $\sim 200$ cm$^{-1}$ with its phonon resonances lying in the mid-IR;[30,31] and GaN-based technology is fully mature, most notably in the micro- and optoelectronics areas, offering a wide range of applications and compatibility with Si integration,[32] the most widely established semiconductor platform. These characteristics taken together position GaN as one of the most promising candidates for the development of plasmonic ENZ devices.

Despite its promise, GaN has been the subject of only a limited number of studies in plasmonics. Although some works have acknowledged its potential as a plasmonic material,[1,17] experimental studies remain scarce and have primarily focused on thick layers (with thicknesses on the order of micrometers),[1,33,34] or more recently on nanostructures such as nanowall networks.[35] However, in-depth analysis of the polariton modes in thin films and demonstration of the plasmonic ENZ mode are still lacking. Thus, the present work provides the first detailed study and experimental demonstration of the plasmonic properties of GaN over Si, exhibiting a plasmonic low-loss ENZ mode in the mid-IR. We report highly doped GaN thin films with a plasma frequency as high as 3350 cm$^{-1}$ ($\sim 3$ μm). By analyzing its optical behavior, we establish GaN as a viable low-loss plasmonic ENZ material in the mid-IR paving the way for its integration into future infrared photonic technologies.

## 2. Results and Discussion

We analyze the plasmonic response of a series of highly doped GaN thin layers in the mid-IR. The films are labeled Samples 1 to 5, with the index increasing with the plasma frequency or carrier concentration (Sample 1 corresponding to the lowest value and Sample 5 to the highest). Each sample consists of a highly doped, 100 nm-thick GaN layer deposited on a 100 nm-thick AlN layer, grown on a Si (111) substrate (see Methods for details). As a reference, an additional sample with the same structure but with an undoped GaN layer is included, referred to as Sample 0.

**Figure 1** shows the measured (solid lines) and simulated (dashed lines) polarized reflectance spectra for all samples, with excellent agreement. The minima between 2500 and 3500 cm$^{-1}$ are clearly visible in the p-polarized reflectance of Samples 1 to 5, which are related to the plasma frequency. Such a minimum is not present in the reflectance spectrum of the undoped Sample 0, as expected. To increase visibility of the plasma-related reflectance dip and its tunability, **Figure 1 (g)** shows the measured reflectance spectra of Samples 1 to 5 in the region where the minimum is observed. Since the plasma frequency, defined as $\omega_p^2 = \frac{Nq}{\epsilon_\infty \epsilon_0 m_{eff}}$ (where $\epsilon_\infty$ is the high energy dielectric constant, $\epsilon_0$ the vacuum permittivity, and q and $m_{eff}$ the electron's charge and effective mass, respectively), is proportional to the electron concentration (N), the shift of the reflectance minimum is indicative of an increasing electron concentration in the thin films.

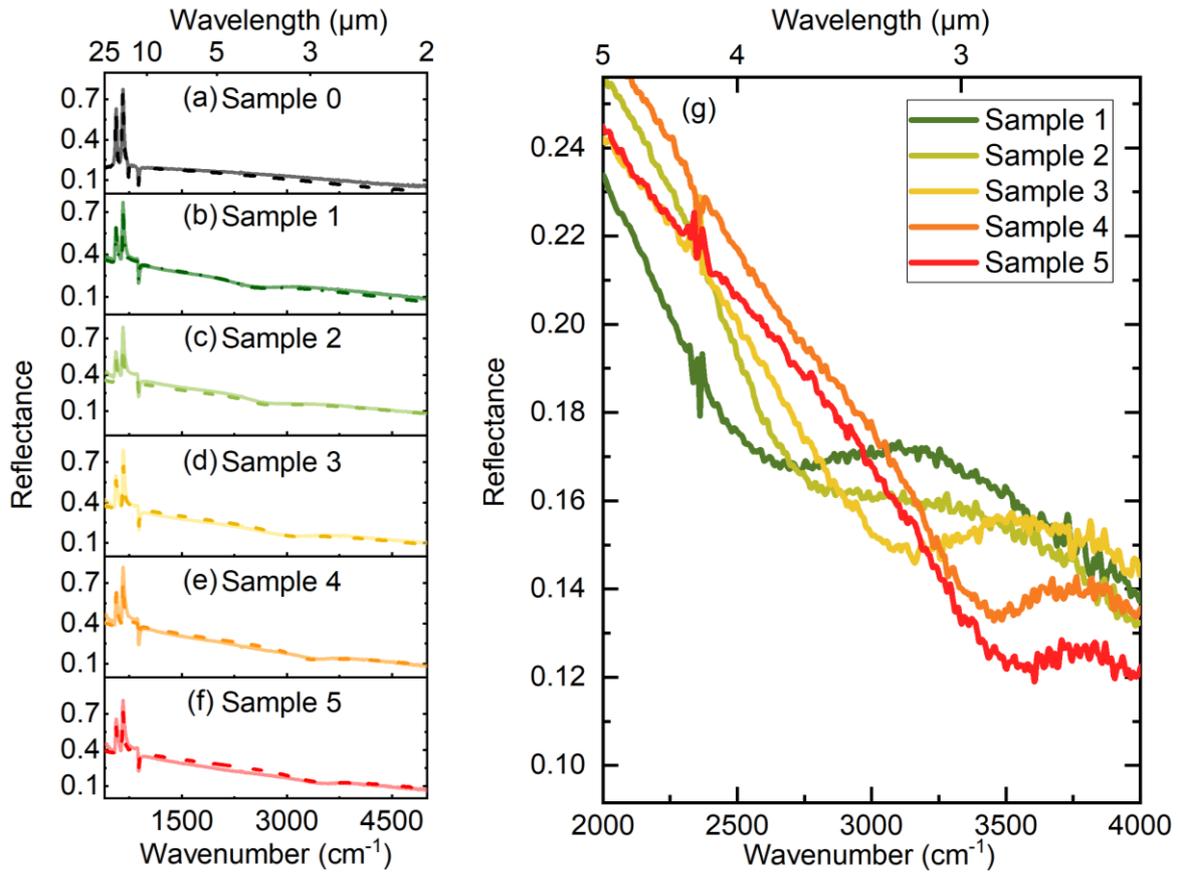

**Figure 1**: (a)-(f) P-polarized reflectance measurements (solid line) and simulated spectra (dashed line) at 45º incidence angle for all samples. (g) Zoomed-in view of the plasma frequency-related minima for Samples 1 to 5.

**Table 1** shows the values of plasma frequency ($\omega_p$), optical losses ($\gamma_p$), electron concentration (N), and ratio of optical losses/plasma frequency, extracted from fitting the measured spectra of all samples (see Methods for details on the measured and simulated p-polarized reflectance spectra). We achieve a tunable plasma frequency ranging from 2500 cm$^{-1}$ (4 μm) to 3350 cm$^{-1}$ (3 μm), corresponding to a change in the electron concentration of the films between 7.4×10$^{19}$ and 1.3×10$^{20}$ cm$^{-3}$. Further, the optical losses in the films are around 25-30% of the plasma frequency in all cases, i.e. independent of the electron concentration. While some works have proposed the potential of GaN as a plasmonic material,[1,17] experimental studies are very limited,[33–35] and highly-doped GaN thin films had not yet been explored in the context of plasmonics. Thus, these results are the first reported experimental optical values of plasma frequency and optical losses in heavily doped GaN thin films.

| Sample | $\omega_p$ (cm$^{-1}$) | $\gamma_p$ (cm$^{-1}$) | N (cm$^{-3}$) | $\frac{\gamma_p}{\omega_p}$ | $\omega[\mathrm{Re}(\epsilon)=0]$ (cm$^{-1}$) | $\mathrm{Im}(\epsilon)|_{\mathrm{Re}(\epsilon)=0}$ |
|--------|------|------|------|------|------|------|
| 1 | 2500 | 810 | 7.4×10$^{19}$ | 30% | 2422 | 1.7 |
| 2 | 2642 | 790 | 8.3×10$^{19}$ | 30% | 2570 | 1.6 |
| 3 | 2988 | 900 | 1.1×10$^{20}$ | 30% | 2895 | 1.6 |
| 4 | 3244 | 820 | 1.2×10$^{20}$ | 25% | 3180 | 1.3 |
| 5 | 3350 | 860 | 1.3×10$^{20}$ | 26% | 3270 | 1.3 |

**Table 1**: Values of the plasma frequency ($\omega_p$), damping ($\gamma_p$), electron concentration ($N$), and ratio of optical losses ($\gamma_p/\omega_p$) of Samples 1 to 5 extracted from the fitting of the measured reflectance spectra. Once those parameters are extracted, the dielectric function of GaN is calculated, and the last two columns present the real part of the dielectric function at the crossover frequency close to $\omega_p$, and the imaginary part of the dielectric function at the same crossover frequency.

These parameters extracted from fitting the measurements, as well as those corresponding to the phonon resonances which will be later discussed in detail, are employed to calculate the dielectric function of GaN for each of the samples. **Figure S1 (a)** in the Supporting information shows that doping the GaN layer causes the zero-crossing point of the real part of the dielectric function associated with the LO phonon resonance ($\omega_{\mathrm{LO,GaN}}$ =

739 cm$^{-1}$) to shift to higher frequencies between $\omega_{LO,GaN}$ and $\omega_p$, due to the hybridization of the plasmonic and phononic resonances. This shift is clearly shown in **Figure S1 (b)** in the Supporting Information, and the values of the zero-crossing point for each sample are shown in **Table 1**. Note that the values of $\omega_p$ and $\omega[Re(\varepsilon) = 0]$ are only equal in the absence of optical losses.

To benchmark our ENZ material against the existing literature in the field, we use the definitions of ENZ and NZI material given in Ref.[1]. The ENZ regime is defined by the condition $|Re(\varepsilon)| < 1$, and the NZI region corresponds to n < 1 (or, equivalently, to a value of $Im(\varepsilon) < 2$ at the crossover frequency). When both conditions are satisfied, the material can be considered a low-loss ENZ material.[1] Additionally, Ref.[1] introduces a figure of merit (FOM) for ENZ materials, defined as the optical losses (given by $Im(\varepsilon)$) at the crossover frequency, and compares this FOM across various materials. In many metals such as titanium, thallium, tungsten and zinc, the crossover frequency lies in the deep ultraviolet owing to their high intrinsic carrier densities, and although they can achieve both the ENZ and NZI conditions, they have not been extensively investigated due to the high frequencies required. Gold and silver are an exception, exhibiting ENZ regions centered at 19000 cm$^{-1}$ (with $Im(\varepsilon)$=3 at the crossover frequency) and at 27500 cm$^{-1}$ ($Im(\varepsilon)$=0.5), respectively. Regarding semiconductors, the best values of the FOM for plasmonic ENZ materials have been reported for highly doped oxides, among which are Ga:ZnO ($Re(\varepsilon)$=0 at 8470 cm$^{-1}$, $Im(\varepsilon)$=0.3), Al:ZnO ($Re(\varepsilon)$=0 at 7700 cm$^{-1}$, $Im(\varepsilon)$=0.3) and Dy:CdO ($Re(\varepsilon)$=0 at 5350 cm$^{-1}$, $Im(\varepsilon)$=0.13). For the particular case of GaN, a predicted value of $Re(\varepsilon)$=0 at ~3 μm and $Im(\varepsilon)$~1 was reported by V. M. Shalaev and A. Boltasseva[17]. To place our material in context, the values of the FOM for all our samples are calculated and listed in the last column of **Table 1**, and result in values of $Im(\varepsilon)$ at the crossover frequency between 1.3 and 1.7. Although these values are slightly higher than those of doped oxides, they compare well with other materials, and all of them satisfy $Im(\varepsilon) < 2$ (equivalent to the condition n < 1 at the crossover frequency),[1] indicating NZI properties and thus low-loss ENZ characteristics. Furthermore, and of particular significance, GaN has the advantage of its integrability with existing micro- and optoelectronic platforms, which is the key factor that makes it an unquestionably exceptional material for active plasmonic applications.

To examine the ENZ and NZI regimes in greater detail, **Figure 2 (a)** shows the calculated dielectric function of GaN for the most heavily doped sample, Sample 5, in the spectral region near the plasma frequency, including the regions where the material satisfies the ENZ and NZI conditions. A spectral band can be observed where these two regimes overlap (from 2990 to 3540 cm$^{-1}$), indicating that the material exhibits low-loss ENZ behavior at those frequencies. This is true for all samples. **Figure 2 (b)** shows the calculated spectral width of these ENZ and NZI regions, for all samples, illustrating the progressive broadening of both regimes with increasing plasma frequency, along with the shift of the crossover frequency shown in **Table 1**. Finally, it is important to emphasize that, although very high doping levels of GaN layers have been demonstrated in thick films,[36] the novelty of our ENZ material resides in achieving similarly high doping levels in thin films (of only 100 nm), which are in good agreement with the values predicted by V. M. Shalaev and A. Boltasseva presented in *Fig.2*. of their review Ref.[1], while being the first experimental in-depth study of GaN from a plasmonics perspective.

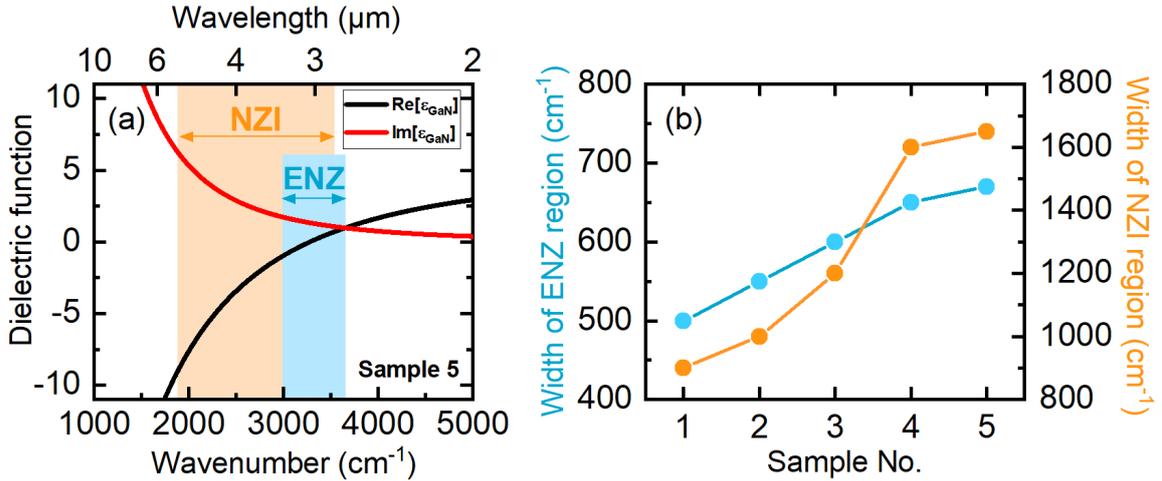

**Figure 2**: (a) Calculated dielectric function of GaN of Sample 5 in the spectral region near the crossover wavelength, including the regions where the material satisfies the ENZ and NZI conditions, in blue and orange, respectively. (b) Blue and orange dots indicate the widths of the ENZ and NZI regions, respectively, for Samples 1 to 5.

Since GaN is a polar material, in addition to holding plasma resonances, it also holds phonon resonances at lower frequencies (longer wavelengths). In order to analyze in detail the phononic region of our structures, **Figure 3 (a)-(f)** shows a zoomed-in view of the spectral region where the phonon frequencies of AlN and GaN are found. Four peaks are visible in the undoped film (Sample 0) corresponding to the TO and LO phonon resonances from AlN ($\omega_{TO,AlN}$ = 661 cm$^{-1}$, $\omega_{LO,AlN}$ = 885 cm$^{-1}$) and GaN ($\omega_{TO,GaN}$ =

565 cm$^{-1}$, $\omega_{LO,GaN}$ = 739 cm$^{-1}$). However, in the doped films the LO resonance from GaN is no longer present **(Figure 3 (b)-(f))** as a result of phonon-plasmon coupling.[33,34,37] As already discussed, the doping of the GaN layer causes the zero-crossing point of the real part of the dielectric function associated with the LO phonon resonance to shift to higher frequencies. However, despite this shift of the zero-crossing point of $\varepsilon_{GaN}$, a very small dip in regular reflectance is still present around 739 cm$^{-1}$ (i.e. around $\omega_{LO,GaN}$ for the undoped sample) in the doped films. This feature is very subtle in regular reflectance measurements and therefore practically indistinguishable in **Figure 3 (a)-(f)**, but becomes clearly observable in ATR configuration, even for the sample with the highest doping level **(Figure 3 (g))**. To explain this observation, we perform calculations of the band structure of GaN thin layers with the electron density of our samples using the BandEng software[38]. Details of the parameters used for the band structure calculations are given in the Methods Section. **Figure 3 (h)** shows the electron concentration (N) as a function of the distance from the surface of Sample 5, revealing the existence of two thin depletion regions – one at each interface of the GaN layer.

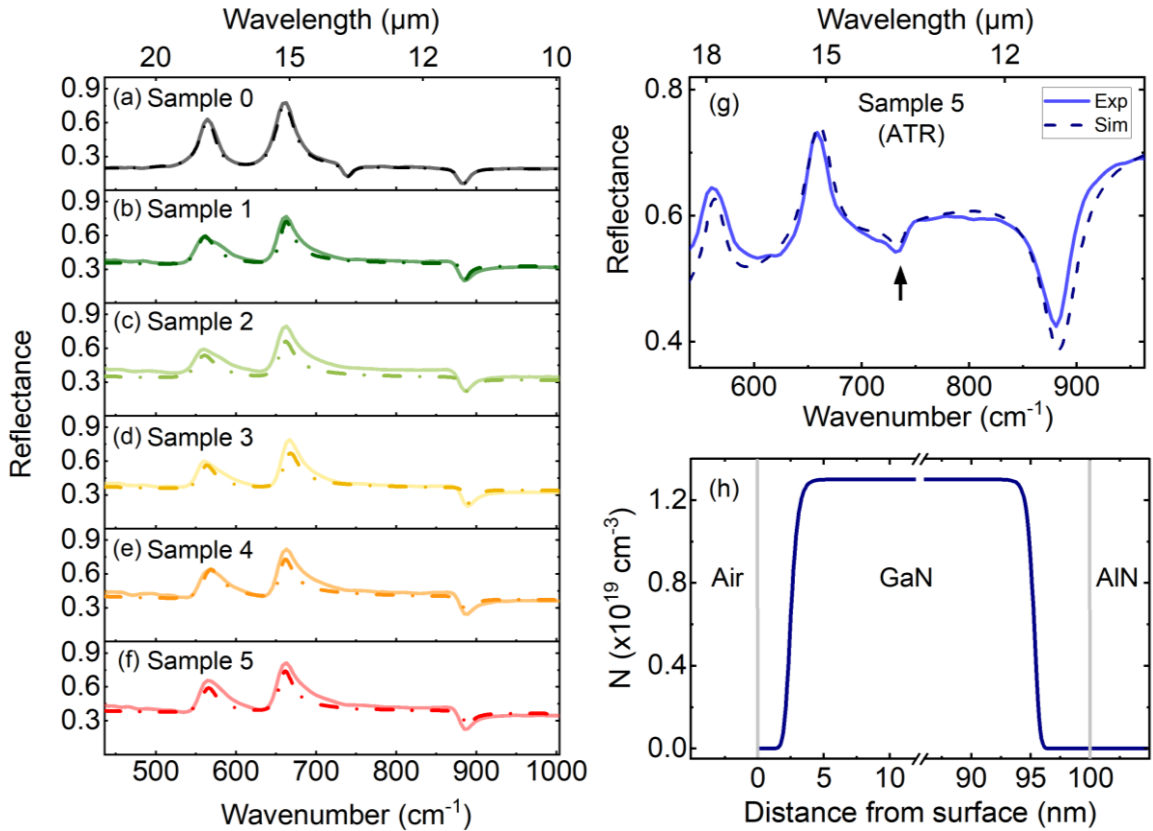

**Figure 3**: (a)-(f) Measured (solid line) and simulated (dashed lines) reflectance spectra for Samples 0 to 5, in the spectral region where the phonon frequencies of AlN and GaN

are found. (g) ATR measurement and model of Sample 5, where a fourth peak (indicated with an arrow) is visible in the phononic region close to $\omega_{LO,GaN} = 739\ cm^{-1}$. (h) Calculation of the electron concentration (N) of Sample 5 as a function of the distance from the surface. Note the break in the X-axis.

The depletion region formed at the Air/GaN interface is approximately 1 nm thick, while the one at the GaN/AlN interface is thicker, of around 4 nm. Although both depletion regions are very thin, together they account for around 5% of the total GaN layer thickness, and therefore, it seems reasonable that they can impact the reflectance spectrum. Indeed, when we include two undoped layers of GaN in the TMM reflectance model of our structure (at the same positions and of thicknesses equal to those of the corresponding depletion regions), the TMM model accurately reproduces the visible feature at 739 $cm^{-1}$, as shown in the fitted spectra in **Figure 3 (g)**. With this simplified method, we can predict the existence of the peak observed at $\omega_{LO,GaN}$, even for the sample with the highest doping level. For completeness, the ATR measurements for all samples are presented in **Figure S2 (a)** in the Supporting Information, where the reflectance dip close to $\omega_{LO,GaN}$ is clearly visible in all of them; and **Figures S2 (b)** and **(c)** show the calculated charge distributions and band structures, respectively, for all samples studied in this work.

Given that the GaN thin layers have a thickness of only 100 nm, we expect the polaritonic mode near the plasma frequency to exhibit ENZ characteristics specific to thin films, namely a resonance that remains non-dispersive with respect to the angle of incidence.[1,4] To demonstrate this experimentally, we performed ATR measurements of Samples 1 and 5 (the least and most doped samples), shown in **Figure 4** ¡Error! No se encuentra el origen de la referencia.**(a), (b)**, respectively. In both cases, the reflectance minima appear to be non-dispersive. To increase visibility, the insets show the measurements after subtracting a straight baseline, clearly revealing that the minimum position does not shift with the angle of incidence. **Figure 4 (c), (d)** shows that the simulated ATR spectra accurately match the experimental reflectance curves for all angles of incidence. We therefore can conclude that our TMM model successfully reproduces the experimental data, and on this basis, we use the parameters extracted from this model to construct the dispersion curves of our samples. These curves are calculated for two main purposes: first, to validate the ENZ nature of the high-energy polariton mode (the one close to $\omega_p$), and second, to carry out a detailed analysis of all polariton modes present in the thin-film structures.

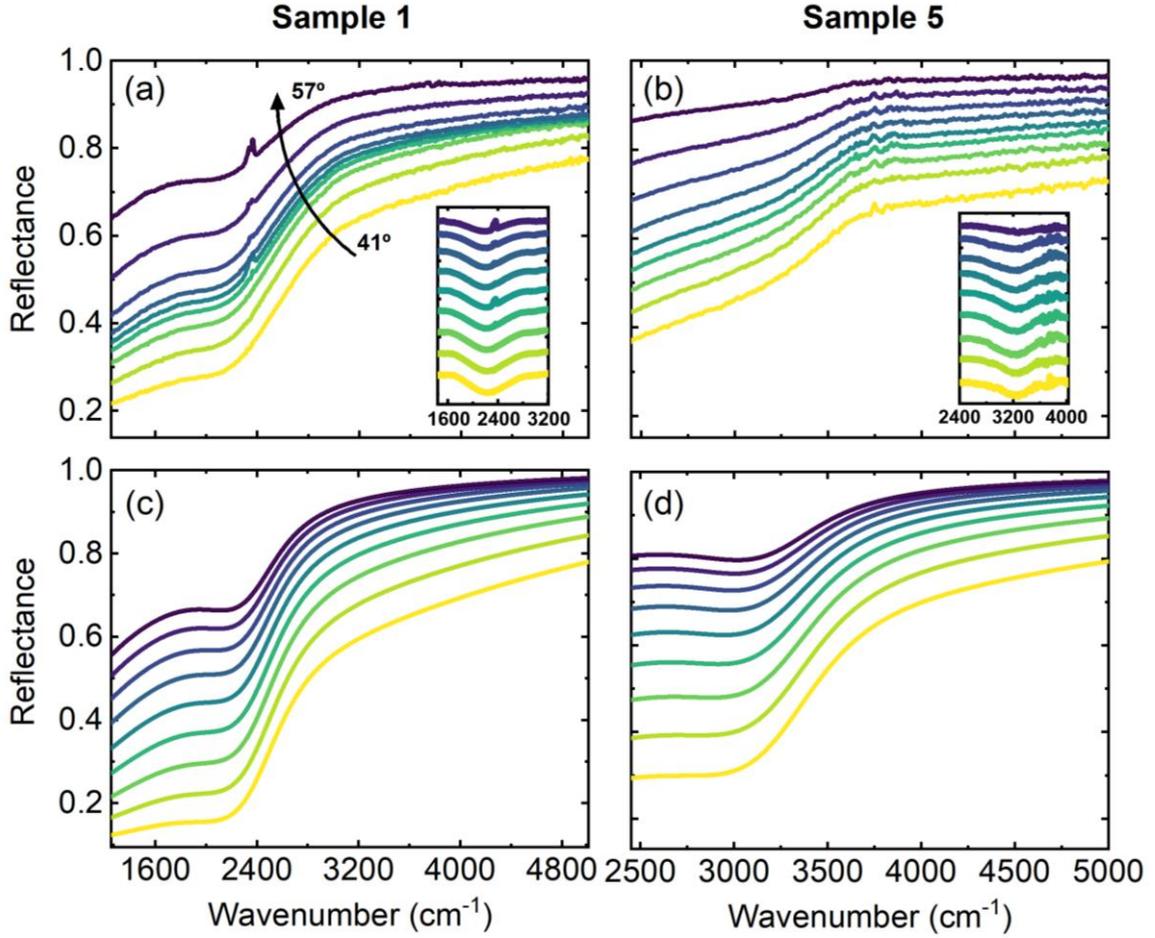

**Figure 4**: (a), (b) ATR angle-resolved measurements and (c), (d) simulated spectra of Samples 1 and 5, respectively. Insets in (a), (b) show the measurements after subtracting a straight baseline, to improve visibility of the minima. The curved arrow in (a) indicates the overall evolution of the spectrum with the angle of incidence, which was varied from 41º to 57º in steps of 2º, for all spectra displayed in this figure.

**Figure 5** shows the calculated dispersion curves of Samples 0 and 5, to analyze the effect that doping the structures has on the hybridization of the modes. The dispersion curves were simulated assuming low optical losses to allow for a clearer visualization. The cone limited by the grey lines represents the measurable range of our experimental ATR setup, and the circular dots are the experimentally resonance frequencies. The asymptotes of all the polariton modes formed at all interfaces are included as horizontal dashed lines and have been calculated as $\epsilon_i = -\epsilon_j$.[39] Thus, the notation used in the figure is "GaN/AlN" for the frequency that satisfies the condition "$\epsilon_{GaN} = -\epsilon_{AlN}$". Since the two depletion regions formed at the GaN layer surfaces do not significantly affect the discussion of the dispersion curves and only add complexity due to the additional number of asymptotes, the dispersion curves were calculated for the simpler structure: Air/100nm GaN/100nm AlN/Si substrate. Nonetheless, for completeness, the dispersion curves for the structure taking into consideration the two depletion regions formed at the GaN layer interfaces is shown in **Figure S3** of the Supporting Information.

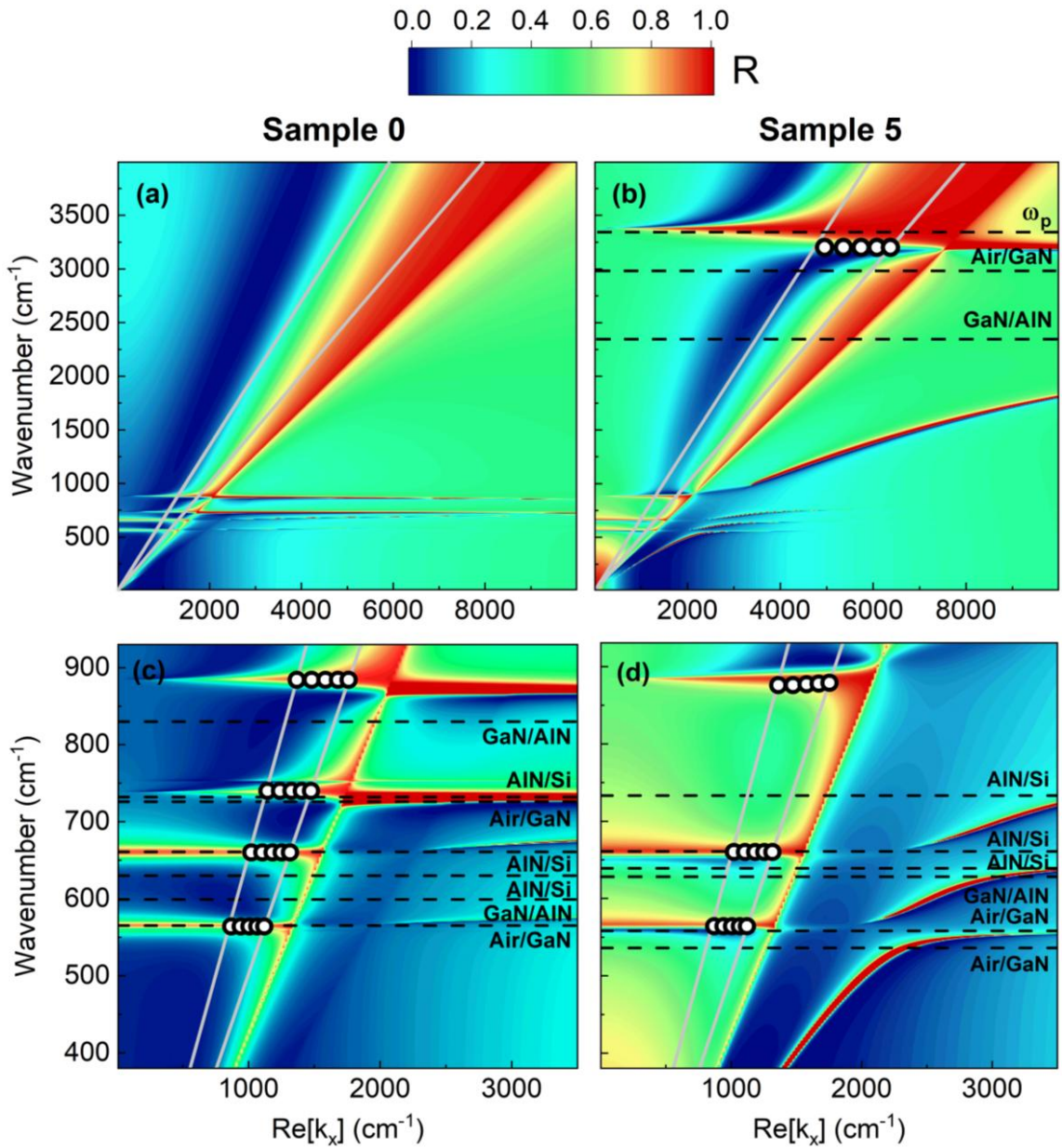

**Figure 5**: Simulated dispersion curves assuming low optical losses for Sample 0, (a) and (c), and for Sample 5, (b) and (d). The cone limited by the grey lines represents the measurable range of our experimental ATR setup and the circular dots are the experimentally determined resonance frequencies. The horizontal dashed lines are the asymptote frequencies of all polariton modes formed at the interfaces.

In order to analyze the dispersion curves in **Figure 5**, we must recall that in a system where a metal film is between two dielectrics with different dielectric functions $\epsilon_1 \neq \epsilon_2$, two surface plasmon polariton (SPP) modes exist, each associated with one of the two distinct dielectric-metal interfaces.[40] These two modes are often referred to as the "high-energy" and "low-energy" SPP modes, which, for instance, in an air-metal-substrate system, correspond to the air-metal and metal-substrate interfaces, respectively. When the thickness of the metal layer is decreased to a value much smaller than the skin depth, the interaction between the two surfaces cannot be neglected and the SPP at one interface "feels" the existence of the SPP at the other interface, giving rise to a "repulsion of levels".[40] This interaction leads to a flat dispersion of the high-energy mode, which becomes pinned at $\omega = \omega_p$, that is associated with an ENZ mode.[4] In the two graphs at the top of **Figure 5**, it can be observed that when the GaN layer of our structures is doped, the expected high- and low-energy modes appear. The high-energy mode is pinned very close to $\omega_p$, as observed experimentally in **Figure 4**, showing a flat dispersion at higher $k_x$ that confirms its ENZ character. The low-energy mode shifts in a slower fashion as $k_x$ is increased. In its way up in frequency, this mode couples to the various surface phonon polariton (SPhP) modes it resonates with, as shown in more detail in **Figure 5 (d)**.

The two graphs at the bottom of **Figure 5** show a zoomed-in view of the phononic region from the top graphs, as to improve visibility of the polaritonic modes in this spectral range. In the undoped GaN case **(Figure 5 (c))**, four non-dispersive modes can be measured within the structure, corresponding to the SPhP modes of the interfaces Air/GaN, GaN/AlN and AlN/Si, approaching the asymptotes indicated in the plot. In the doped GaN case **(Figure 5 (d))** the modal behavior undergoes huge changes. Firstly, the mode present at approx. 739 cm$^{-1}$ in the undoped sample no longer exists due to the shift of the zero-crossing point of $\varepsilon_{GaN}$, already discussed. Secondly, the remaining three modes are still visible at similar frequencies experimentally, but their behavior changes completely, approaching new asymptotes due to the modification of the GaN dielectric function by the additional Drude term. In particular, the change in the dispersion of the mode near 900 cm$^{-1}$ can be observed experimentally. **Figure S4** of the Supporting

Information shows ATR measurements of Samples 0 and 5, illustrating how the resonance frequency of this mode becomes much more dispersive due to its hybridization with the SPP modes.

To sum up, this thorough experimental analysis serves as the first demonstration of a plasmonic ENZ mode in GaN thin films, while also providing a detailed discussion of how the coupling of the surface plasmon and phonon polariton modes affects the dispersion curves of these structures.

## 3. Conclusions

In this work we have demonstrated heavily doped GaN as an ENZ material. To do so, we have used thin films deposited over AlN/Si that have plasma frequencies between 2500 $cm^{-1}$ (4 µm) and 3350 $cm^{-1}$ (3 µm) and optical losses in the 25-30 % range, corresponding to electron concentrations from $7.4 \times 10^{19}$ to $1.3 \times 10^{20}$ $cm^{-3}$. Taking together all previous results, we are able to build the dielectric function of the heavily doped GaN layers, where we can identify both the ENZ and NZI regions, which define the low-loss ENZ regime when they overlap. In the most heavily doped sample with the largest plasma frequency, the low-loss ENZ region extends roughly from 2990 $cm^{-1}$ (~3.3 µm) to 3540 $cm^{-1}$ (~2.8 µm), opening the possibility of integrating plasmonic layers featuring strong sub-wavelength confinement characteristics with widely used GaN micro- and optoelectronic technologies.

Using a combination of ATR measurements and TMM modeling, we have analyzed in detail the complex family of polaritons that arise in the structures. As it is characteristic of thin films, the two high- and low-energy branches of the surface plasmon polaritons are observed, which are heavily affected by the presence of phonons in both the highly polar GaN and AlN layers. Hence, the resulting surface polaritons have a hybrid plasmonic-phononic character, as it is clearly observed in the dispersion curves in a high frequency range. An analysis of the high-energy SPP branch in GaN, found at a frequency close to the plasma frequency, indicates that it has a flat dispersion with the in-plane momentum, thereby demonstrating it is a plasmonic ENZ mode.

## 4. Methods

***Sample growth:*** All the samples employed in this work were grown by molecular beam epitaxy (MBE) on resistive Si (111) substrates. Each sample consists of a 100-nm Ge:GaN layer deposited on a 100-nm AlN layer, grown on a Si (111) 2-inch substrate. The AlN buffer layer was grown following the method outlined in Ref.[41]. Ga, Ge, and Al elements are supplied by Knudsen cells, while N is produced through on-surface cracking of ammonia. We previously showed that high electron densities can be achieved in Ge-doped GaN or AlGaN layers suitable for tunnel junctions.[42] The Ge:GaN layers are grown at a rate of 10 nm/min and a temperature range of 740-760°C. The doping level depends on the Ge cell temperature, which ranges from 1050°C to 1200°C. The growth front tends to become rough due to very high Ge doping. To limit this roughening, two annealing steps are performed at 810°C. This enables the production of layers with a root-mean-square (RMS) roughness of 1.5-1.8 nm for atomic force microscopy scans of 10x10 μm². In comparison, the undoped sample has an RMS roughness of 0.6 nm.

***Polarized reflectance spectroscopy measurements and ATR measurements:*** Reflectance measurements were carried out in a Fourier Transform Infrared (FTIR) Spectrometer, using *p*-polarized incident light. Additionally, to match the light in-plane momentum to the polariton momentum and excite the SPPs and SPhPs, the attenuated total reflectance (ATR) technique has been used in Otto configuration, with a ZnSe prism ($n_{prism} = 2.37$). The angle of incidence (θ) at the prism−air gap interface was controlled from 39° to 59°, allowing to scan the polariton dispersion curve through $k_x = k_0 n_{prism} \sin(\theta)$, where $k_0$ is the light momentum in vacuum.

***Numerical models:*** *(i) TMM model with full layer structure:* To define the structure, we account for all materials, in this case: GaN, AlN and Si, each of them defined by its individual dielectric function and thickness. To consider the depletion regions formed at both interfaces of the GaN layer, we add two undoped GaN layers to the structure, so that the final layer structure we simulate in the model is: Air/depletion region/GaN/depletion region/AlN/Si substrate. To obtain the simulated reflectance spectra of our structures and to fit the model to the experimental measurements to obtain the parameters of the measured sample, we employ a commercial package based on the transfer matrix method (TMM). *(ii) Dispersion curves:* We calculate the dispersion curves with an in-house TMM-based package.

*Calculations of the band structure and charge distribution:* Self-consistent Poisson– Schrödinger simulation software, BandEng,[38] was used to determine the band structure and charge distribution of our structures. The software takes all the necessary material properties as inputs, including the materials band gap, the electron affinity and the effective masses of carriers in the material, as well as polarization properties such as spontaneous polarization, the elastic and piezoelectric constants. All these parameters are well known and have been extracted from the literature.[43,44] The software also allows to set a barrier height at the interface of the heterostructure. In our case, the surface potential at the Air/GaN interface was set to 0.6 eV, consistent with several reports in the literature indicating typical values in the range of 0.5 to 0.8 eV.[44,45] Nevertheless, small variations in this value have a negligible effect on the results.

## Supporting Information

Supporting Information is available from the Wiley Online Library or from the author.

## Acknowledgements

This work was partly supported by Projects PID2024-156706OB-C2 and PDC2023-145827-C2 funded by MICIU/AEI/10.13039/501100011033/. The authors also acknowledge the support provided by the MICRONANOFABS ICTS network from the MICIU.

## Data Availability Statement

The data that support the findings of this study are available from the corresponding author upon reasonable request.

**Table of Contents**